\documentclass[twocolumn,10pt,aps,pra,amsfonts,amsmath,amssymb,superscriptaddress]{revtex4-1}

\usepackage{graphicx}
\usepackage{times}
\usepackage[USenglish]{babel}
\usepackage[hidelinks,implicit=false,pdfpagemode=UseNone]{hyperref}

\newcommand{\LMU}{Fakult\"at f\"ur Physik, Munich Quantum Center, and Center for NanoScience~(CeNS), Ludwig-Maximilians-Universit\"at M\"unchen, Geschwister-Scholl-Platz 1, 80539 M\"unchen, Germany}
\newcommand{\LANL}{Los Alamos National Laboratory (LANL), Los Alamos, New Mexico 87545, USA}
\newcommand{\Rice}{Department of Chemical and Biomolecular Engineering, Rice University, Houston, Texas 77005, USA}
\newcommand{\corresponding}{andre.neumann@lmu.de; alexander.hoegele@lmu.de}

\begin{document}

\title{Signatures of defect-localized charged excitons in the photoluminescence of monolayer molybdenum disulfide}

\author{Andre Neumann}\email{\corresponding}\affiliation{\LMU}
\author{Jessica Lindlau}\affiliation{\LMU}
\author{Manuel Nutz}\affiliation{\LMU}
\author{Aditya D. Mohite}\affiliation{\Rice}
\author{Hisato Yamaguchi}\affiliation{\LANL}
\author{Alexander H\"ogele}\email{\corresponding}\affiliation{\LMU}

\begin{abstract}
We study spatial photoluminescence characteristics of neutral and charged excitons across extended monolayer MoS$_2$ synthesized by chemical vapor deposition. Using two-dimensional hyperspectral photoluminescence mapping at cryogenic temperatures we identify regions with increased emission from charged excitons associated with both spin-orbit split valence subbands. Such regions are attributed to unintentional doping at defect sites with excess charge that bind neutral excitons to form defect-pinned trions. Our findings imply comparable timescales for the formation, relaxation, and radiative decay of $B$ trions, and add defect-localized $A$ and $B$ trions to the realm of photoexcited quasiparticles in layered semiconductors.

\bigskip
\noindent Published in Phys. Rev. Materials \textbf{2}, 124003 (2018). Copyright 2018 American Physical Society.

\noindent The version of record is available online at \url{https://doi.org/10.1103/PhysRevMaterials.2.124003}.
\end{abstract}

\maketitle

The photophysics of semiconducting transition metal dichalcogenide (TMD) monolayers (MLs) such as molybdenum disulfide (MoS$_2$) are dominated by excitonic phenomena. The inherently two-dimensional material properties with strong confinement in space and reduced dielectric screening lead to tightly bound electron-hole pairs with binding energies exceeding several hundred of meV \cite{He2014, Chernikov2014, Ye2014}. The two lowest band-edge excitons, commonly referred to as $A$ and $B$ excitons, originate from the spin-orbit split valence subbands at the $K$ and $K'$ valleys of the hexagonal Brillouin zone \cite{Splendiani2010, Mak2010}. Due to strong spin-orbit effects and broken inversion symmetry in real space, the exciton spins are pinned to their valley degree of freedom and can be addressed selectively or coherently with polarized light \cite{Mak2012, Zeng2012, Jones2013}, enabling their manipulation for valleytronic applications \cite{Xu2014, Schaibley2016}. Furthermore, large Coulomb interactions in TMDs give rise to pronounced many-body phenomena where quasiparticles consisting of three or more charge carriers are formed, including trions \cite{Mak2013, Ross2013, Jones2013}, biexcitons \cite{You2015, Sie2015, Hao2017}, or Fermi polarons \cite{Sidler2017}.

The kinetics responsible for the photogeneration of stable many-body systems are complex \cite{Almand-Hunter2014}. For example, in the limit of low densities of excess charge carriers, trions may form through coalescence of an exciton with an extra charge within the Fermi sea or pinned to a defect, or evolve from an electron-hole plasma in a nonequilibrium state \cite{Portella-Oberli2009}. In most cases the number of photogenerated trions is determined by the former channel and depends on the doping level of the semiconductor.

In our work we studied electrically and chemically doped ML MoS$_2$ with confocal photoluminescence (PL) spectroscopy at cryogenic temperatures. ML MoS$_2$ crystals were grown by chemical vapor deposition (CVD) as described previously \cite{Najmaei2013, Neumann2017}. After synthesis, the flakes were transferred onto substrates of highly $p$-doped silicon (Si) with $100$~nm of thermal silicon dioxide (SiO$_2$). A field-effect transistor (FET) was fabricated for electrostatic doping of one sample. On the second sample grown under identical conditions we studied bare MoS$_2$ flakes without electric contacts yet with local chemical doping effects due to unintentionally formed point defects \cite{Hong2015} which promote adsorption of molecular contaminants \cite{Lin2016, Cai2016}. In the experiments described below we used the FET sample to calibrate the charge polarity of the background doping characteristic of ML MoS$_2$ on SiO$_2$ \cite{Radisavljevic2011, Najmaei2013} and the associated spectral signatures of $A$ and $A^-$ PL stemming from neutral and negatively charged excitons of the $A$ exciton manifold \cite{Mak2013, Tongay2013a, Mouri2013, Cadiz2016}. In general, the two PL resonances can be ascribed to many-body generalizations of trion bound and unbound states or excitons dressed by the Fermi sea, and represent trions and excitons at sufficiently low carrier densities \cite{Efimkin2017}. Based on the reference measurements of the first sample, we used the second sample to study the PL characteristics of neutral and charged excitons of both $A$ and $B$ spin-orbit split manifolds. Our results identify the emission from defect-localized negatively charged trions, $B^-$, as excited-state counterparts of the charged $A$ exciton. Moreover, from the analysis of the associated charge doping profiles obtained in PL imaging we provide insight into the trion formation dynamics for both exciton manifolds.

Our cryogenic PL spectroscopy and raster-scan imaging experiments were performed in a helium bath cryostat at $4.2$~K or a closed-cycle cryostat with a base temperature of $3.1$~K. The samples were positioned into the focal plane of a low-temperature apochromatic objective with a numerical aperture of $0.65$ and diffraction-limited confocal excitation and detection spots of $\sim 0.7~\mu$m diameter. PL and Raman measurements were performed with lasers at $532$~nm or $639$~nm and a spectrometer equipped with a nitrogen-cooled Si CCD. The spectral resolution of the system was $\sim 0.35$~meV for PL and $\sim 0.6$~cm$^{-1}$ for Raman spectroscopy.

\begin{figure}[t]
\begin{center}
\includegraphics[scale=1.015]{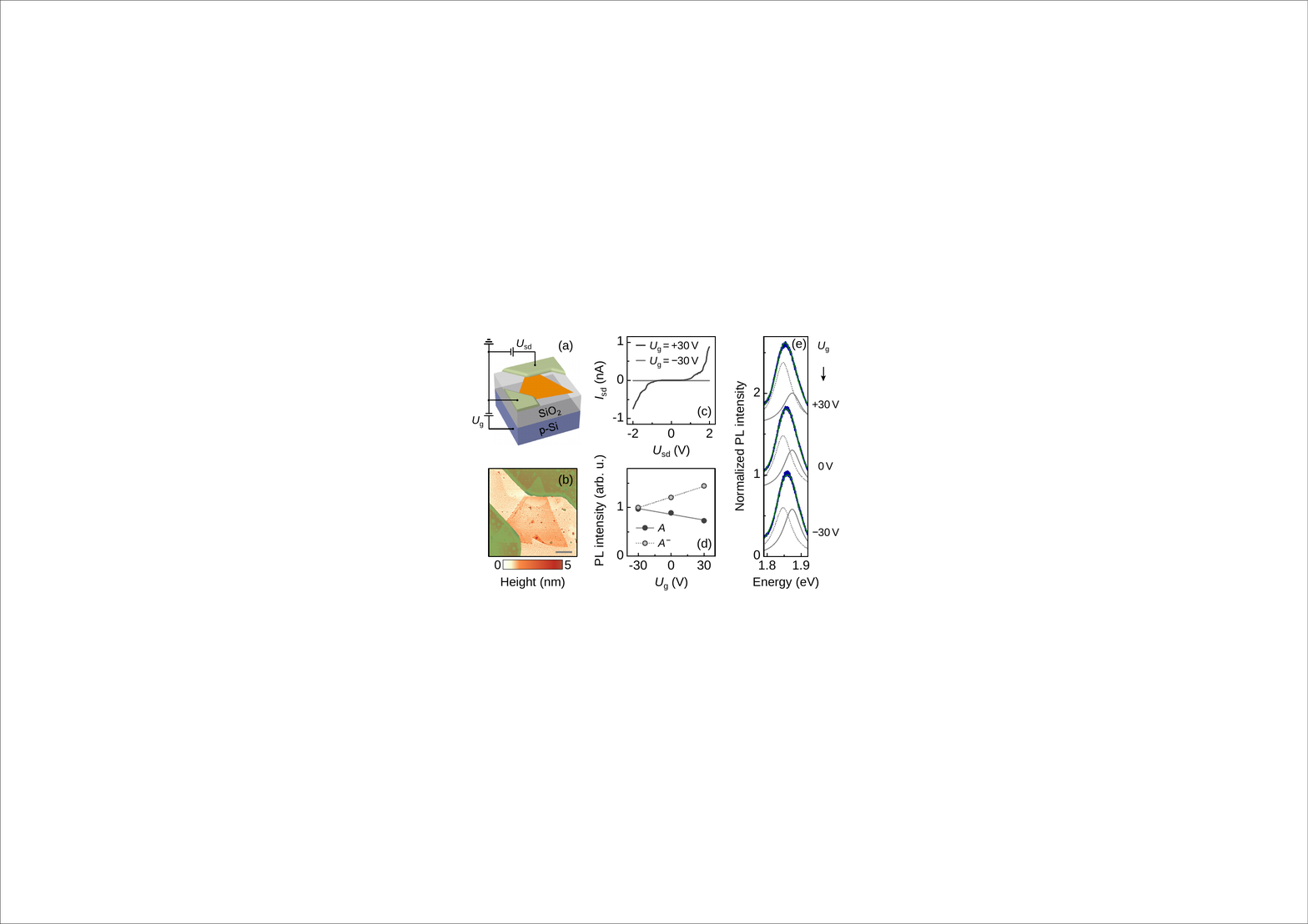}
\caption{(a) Schematic of a field-effect transistor based on monolayer MoS$_2$ (orange triangle) with source and drain contacts (green-colored metallic pads) on top of $100$~nm thermal SiO$_2$ and $p$-doped Si back gate (dark blue). (b) Atomic force micrograph of the device (scale bar is $5~\mu$m). (c) Source-drain current ($I_\mathrm{sd}$) through the flake as a function source-drain voltage ($U_\mathrm{sd}$) for two gate voltages ($U_\mathrm{g}$) of $\pm 30$~V. (d) Integrated $A$ and $A^-$ photoluminescence intensities as a function of $U_\mathrm{g}$ under excitation at $532$~nm (solid and dotted lines are linear guides to the eye). (e) Corresponding photoluminescence spectra (blue) with double-Lorentzian fits (green) accounting for $A$ (solid gray) and $A^-$ (dotted gray) contributions; the spectra were offset for clarity. Data in (c)--(e) were measured at $4.2$~K.}
\label{fig1}
\end{center}
\end{figure}

First, we studied the effect of intentional doping of MoS$_2$ ML flakes obtained with our sample fabrication method involving CVD and transfer to target substrates \cite{Reina2008}. The schematics and an atomic force micrograph of our ML MoS$_2$ FET device are shown in Figs.~\ref{fig1}(a) and \ref{fig1}(b), respectively. Electrical contacts on top of a MoS$_2$ ML triangle were defined by means of laser lithography and deposition of titanium ($3$~nm) and gold ($50$~nm). The $p$-doped Si substrate was used as the back gate of the device. The FET exhibited $n$-type conductivity in cryogenic current-voltage measurements [Fig.~\ref{fig1}(c)] with non-Ohmic response of the source-drain current that is characteristic of non-annealed devices with sizable Schottky barriers \cite{Radisavljevic2011, Buscema2013, Yamaguchi2015}.

%%%%%%%%%%%%%%%%%%%%%%%%%%%%%%%%%%%%%%%%%% FIG 1 %%%%%%%%%%%%%%%%%%%%%%%%%%%%%%%%%%%%%%%%%%
%%%%%%%%%%%%%%%%%%%%%%%%%%%%%%%%%%%%%%%%%%%%%%%%%%%%%%%%%%%%%%%%%%%%%%%%%%%%%%%%%%%%%%%%%%%

The tuning of the doping profile in the low carrier density regime with negligible screening effects was also monitored with PL spectroscopy. For MoS$_2$ MLs on SiO$_2$ both neutral $A$ excitons and negative $A^-$ trions with a splitting of $\sim 30$~meV, corresponding to the trion binding energy, contribute to the emission spectrum \cite{Mak2013, Cadiz2016}. In Fig.~\ref{fig1}(d) we plot the $A$ and $A^-$ PL intensities as a function of the applied gate voltage. The data points were obtained from spectral deconvolution of the total PL spectrum into the emission from $A$ and $A^-$ excitons, each modeled by a Lorentzian with full-width at half-maximum (FWHM) linewidth of $60$~meV and peak maxima separated by the trion binding energy [Fig.~\ref{fig1}(e); note the correspondence between the spectrum in blue and the model fit in green]. For zero gate bias the imbalanced intensities of $A$ and $A^-$ correspond to an electron doping density on the order of $10^{12}$~cm$^{-2}$ \cite{Lee2014} due to substrate-induced background doping of MoS$_2$ on SiO$_2$ \cite{Dolui2013, Scheuschner2014, Kang2017}. An increase in the electron concentration at a gate voltage of $+30$~V was accompanied by an increase of $A^-$ and a decrease of $A$ emission and the opposite effect was observed at $-30$~V [Fig.~\ref{fig1}(d)] in accord with a partial compensation of the negative background doping at negative gate voltages.

%%%%%%%%%%%%%%%%%%%%%%%%%%%%%%%%%%%%%%%%%% FIG 2 %%%%%%%%%%%%%%%%%%%%%%%%%%%%%%%%%%%%%%%%%%
\begin{figure}[t]
\begin{center}
\includegraphics[scale=1.015]{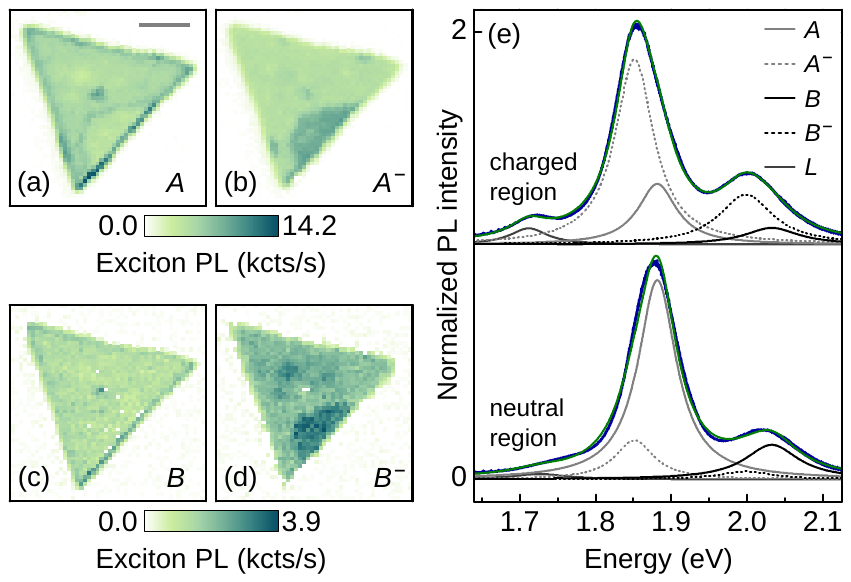}
\caption{(a) and (b) Raster-scan maps of $A$ and $A^-$ photoluminescence intensities in a monolayer MoS$_2$ crystal with a region of defect-induced chemical doping at the lower right edge (scale bar is $5~\mu$m). (c) and (d) Same for $B$ and $B^-$ emission. (e) Normalized photoluminescence spectra measured on charged and neutral positions of the crystal (top and bottom blue traces, respectively) with Lorentzian fits of neutral and negatively charged $A$ and $B$ exciton emission as well as localized excitons ($L$) below $\sim 1.75$~eV (gray traces with inset labels); green traces are sums of all five Lorentzians. The images in (a)--(d) were constructed by integrating the respective contributions of $A$, $A^-$, $B$, and $B^-$ emission to the photoluminescence spectra at each raster-scan position (pixels with signal-to-noise ratios below $10$ were set to zero). All data were measured at $3.1$~K with $532$~nm excitation.}
\label{fig2}
\end{center}
\end{figure}
%%%%%%%%%%%%%%%%%%%%%%%%%%%%%%%%%%%%%%%%%%%%%%%%%%%%%%%%%%%%%%%%%%%%%%%%%%%%%%%%%%%%%%%%%%%

Complementary to field-effect doping, chemical doping provides an alternative approach to trion generation in ML TMDs \cite{Mouri2013, Peimyoo2014, Lui2014}. Molecular adsorption and defect sites in layered semiconductors give rise to binding of excitons to immobile charges \cite{Tongay2013, Nan2014}, thereby forming defect-localized trions. In this regime of localized excitonic complexes \cite{Godde2016}, we expand our analysis to a previously studied MoS$_2$ single crystal \cite{Neumann2017} without active charge control from the same CVD batch as the sample of Fig.~\ref{fig1}. By raster scanning the sample with respect to diffraction-limited confocal excitation and detection spots, PL spectra spanning the range of $A$ and $B$ excitons were recorded. At each pixel of the raster scans the spectra were decomposed into five Lorentzians. Four distributions with fixed energies and linewidths were used to fit $A$ and $B$ excitons and their trion counterparts $A^-$ and $B^-$ \cite{Berkelbach2013}, and lower energy localized $L$ excitons were accounted for by an additional Lorentzian with variable peak energy and linewidth parameters. These conditions restrict the validity of our evaluation to low charge doping of $\lesssim 10^{12}$~cm$^{-2}$, where the exciton-trion peak difference is approximately constant \cite{Chernikov2015a, Courtade2017} and the picture of dressed exciton states is not relevant \cite{Efimkin2017}. Results for $A$ and $B$ excitons and trions are shown in Figs.~\ref{fig2}(a)--\ref{fig2}(d). The observation of $B$ excitons in PL indicates that relaxation pathways from the $B$ to the $A$ exciton manifold which require spin or valley flips of both electrons and holes \cite{Xiao2012} are slow as compared to the radiative $B$ exciton recombination time.

For direct comparison, Fig.~\ref{fig2}(e) shows representative spectra exhibiting mostly neutral and charged excitons in the MoS$_2$ ML (note the increased emission from localized $L$ excitons in the trion-dominated spectrum suggesting local correlations between trion peak intensity and disorder). Best fits for $A$ excitons at $1.88$~eV and $B$ excitons at $2.03$~eV were obtained with inhomogeneously broadened FWHM linewidths of $60$~meV and $85$~meV, respectively, and the same linewidths for the corresponding trions. Our analysis verifies that the $B^-$ negatively charged trion counterpart to $A^-$ is also stable \cite{Berkelbach2013} with a slightly larger trion binding energy of $34$~meV. The small difference in the trion binding suggests comparable extends of the Bohr radii for $A^-$ and $B^-$ trions, consistent with diamagnetic shifts of neutral $A$ and $B$ excitons in the related ML material WS$_2$ \cite{Stier2016}. Our findings of luminescent $B^-$ trions complement their observation in reflectance contrast measurements in ML WS$_2$ \cite{Chernikov2015a} as well as in photoinduced absorption of few-layer MoS$_2$ \cite{Borzda2014}.

%%%%%%%%%%%%%%%%%%%%%%%%%%%%%%%%%%%%%%%%%% FIG 3 %%%%%%%%%%%%%%%%%%%%%%%%%%%%%%%%%%%%%%%%%%
\begin{figure}[t!]
\begin{center}
\includegraphics[scale=1.015]{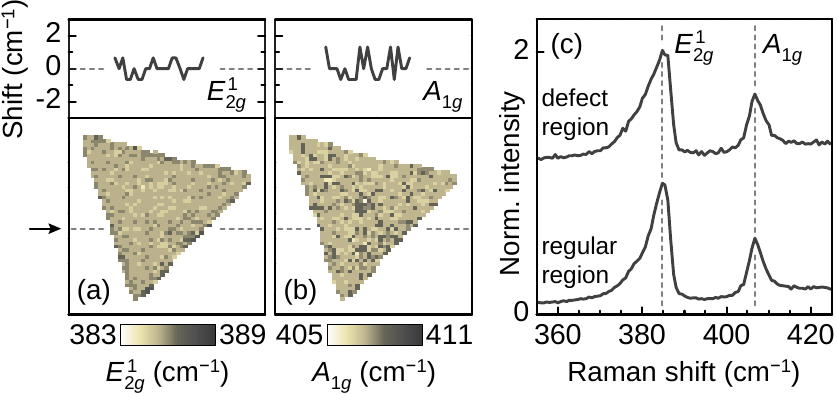}
\caption{(a) and (b) Raster-scan maps of the $E^{1}_{2g}$ and $A_{1g}$ Raman mode frequencies (bottom panels) and line cuts along the arrow (top panels) recorded with an excitation laser at $639$~nm on the same MoS$_2$ crystal as in Fig.~\ref{fig2}. The line cuts are shown as frequency shifts relative to $385$~cm$^{-1}$ and $407$~cm$^{-1}$ for the $E^{1}_{2g}$ and $A_{1g}$ modes, respectively. (c) Representative Raman spectra of the $E^{1}_{2g}$ and $A_{1g}$ modes for regular (bottom) and defect-contaminated (top) positions of the MoS$_2$ crystal (spectra were normalized and offset for clarity; dashed lines are guides to the eye). The mode assignment corresponds to common bulk notation in MoS$_2$ at the $\Gamma$ point of the Brillouin zone. All data were measured at $3.1$~K.}
\label{fig3}
\end{center}
\end{figure}
%%%%%%%%%%%%%%%%%%%%%%%%%%%%%%%%%%%%%%%%%%%%%%%%%%%%%%%%%%%%%%%%%%%%%%%%%%%%%%%%%%%%%%%%%%%

Apart from charge doping, ML spectra are also known to be sensitive to strain \cite{He2013, Conley2013, Zhu2013}. To eliminate any related ambiguity in the origin of spectral shifts of $A$ and $B$-type excitons, we inspected the crystal structure of the MoS$_2$ flake in Fig.~\ref{fig2} by means of Raman spectroscopy. Vibrational in-plane modes in TMDs such as the $E^{1}_{2g}$ mode are most affected by strain. In MLs the Raman shift of the $E^{1}_{2g}$ mode is known to decrease by $2.1$~cm$^{-1}$ per percent of uniaxial strain \cite{Rice2013} without dependence on charge doping \cite{Chakraborty2012}. For our sample the homogenous profile of the $E^{1}_{2g}$ mode around $385$~cm$^{-1}$ shown in Fig.~\ref{fig3}(a) suggests strain variations across the crystal of at most $\sim 0.3$\%. These variations can only account for exciton redshifts of less than half of the measured binding energies of $A^-$ and $B^-$ trions. Opposed to $E^{1}_{2g}$ vibrations, the frequency of the out-of-plane $A_{1g}$ mode is not sensitive to uniaxial strain \cite{Rice2013} but softens from charge neutrality to an electron doping density of $\sim 2 \times 10^{13}$~cm$^{-2}$ by $4$~cm$^{-1}$ \cite{Chakraborty2012}. Fig.~\ref{fig3}(b) shows the spatial distribution of the $A_{1g}$ frequency centered at $407$~cm$^{-1}$ for our ML crystal. The profile reveals uniformity within $\sim 2$~cm$^{-1}$ and an average $A_{1g}$ to $E^{1}_{2g}$ peak separation of $\sim 22$~cm$^{-1}$ that is typical for CVD-grown MoS$_2$ MLs at cryogenic temperatures \cite{Lanzillo2013}. With this data at hand, we confirm that the flake exhibited only low doping well below the electron concentration of $10^{13}$~cm$^{-2}$ with local changes bordering the sensitivity of our experiment. In fact, high signal-to-noise Raman spectra of the $A_{1g}$ and $E^{1}_{2g}$ modes at representative chemically doped and regular ML positions of the flake were identical within our resolution limit [Fig.~\ref{fig3}(c)]. We therefore conclude that the MoS$_2$ ML is weakly and locally doped without substantial strain changes.

%%%%%%%%%%%%%%%%%%%%%%%%%%%%%%%%%%%%%%%%%% FIG 4 %%%%%%%%%%%%%%%%%%%%%%%%%%%%%%%%%%%%%%%%%%
\begin{figure}[t!]
\begin{center}
\includegraphics[scale=1.015]{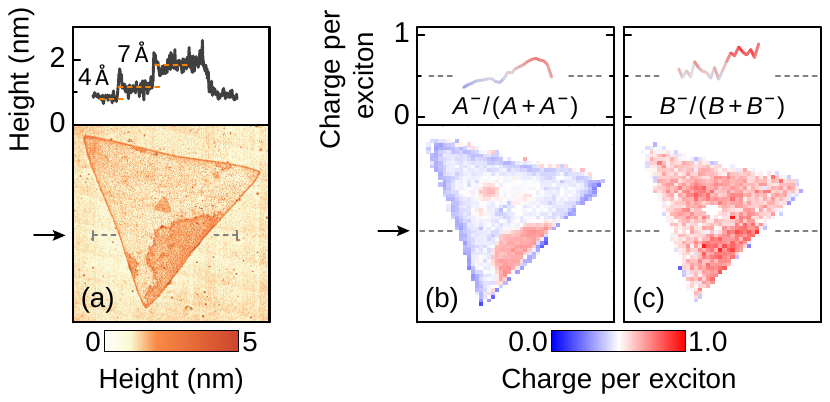}
\caption{(a) Atomic force surface topography for the MoS$_2$ flake of Fig.~\ref{fig2}: micrograph (bottom panel) and height profile across the dashed line (top panel). The step heights of $\sim 4$~{\AA} and $\sim 7$~{\AA} identify regular and defective monolayer regions, respectively. (b) Map of charge doping in units of electron charge per $A$ exciton (bottom) for the same flake and corresponding profile along the line indicated by the arrow (top). The map was computed as $A^-/(A+A^-)$, where $A$ ($A^-$) are the deconvolved exciton (trion) photoluminescence intensities. (c) Equivalent map (bottom) and doping profile (top) for $B$ excitons, computed as $B^-/(B+B^-)$ with $B$ and $B^-$ contributions obtained from spectral deconvolution. Photoluminescence data were acquired at $3.1$~K with $532$~nm excitation; pixels with signal-to-noise ratios below $40$ in (b) and $20$ in (c) were omitted.}
\label{fig4}
\end{center}
\end{figure}
%%%%%%%%%%%%%%%%%%%%%%%%%%%%%%%%%%%%%%%%%%%%%%%%%%%%%%%%%%%%%%%%%%%%%%%%%%%%%%%%%%%%%%%%%%%

Further support in favor of defect-pinned doping was obtained by mapping the surface topography of the crystal [Fig.~\ref{fig4}(a)]. Apart from regular ML domains identified by the step of $\sim 4$~{\AA} above the substrate, our measurements clearly show regions elevated by $\sim 7$~{\AA} on top of the ML, such as the trion-rich puddle at the lower edge. We assume that these elevated regions stem from accumulations of molecular adsorbates at negatively charged sulfur vacancy sites \cite{Qiu2013}, which are commonly encountered in CVD-grown MoS$_2$ due to their low formation energy \cite{Zhou2013}. The correspondence in the observation of $A^-$ and $B^-$ trions at the same positions implies that the localized electronic charges of sulfur vacancies were not fully screened by the molecules acting as acceptors.

Finally, we qualitatively discuss local charge inhomogeneities in the chemically doped MoS$_2$ ML using the ratio of densities of charged to neutral excitons. To this end, we compute charge doping profiles as $A^-/(A+A^-)$ for the ground state excitons and $B^-/(B+B^-)$ for the spin-orbit split excitons [Figs.~\ref{fig4}(b) and \ref{fig4}(c), respectively]. The profiles highlight in red (blue) local regions with higher (lower) electron concentrations, among them the surface-contaminated puddle. At regular positions away from the puddle and edges, the charge concentration of $\sim 0.5$ electrons per $A$-type exciton [Fig.~\ref{fig4}(b)] is found in good agreement with the background doping of our electrostatically tunable ML at zero gate voltage [Fig.~\ref{fig1}(d)]. In the limit of low doping, and assuming trion formation through exciton-electron Coulomb interactions on timescales $\tau_\mathrm{f}$ fast compared to the exciton decay time $\tau$, the profiles provide a quantitative means to image the local doping level by employing the mass action model \cite{Ross2013, Mouri2013, Peimyoo2014}. However, considering that PL decay typically occurs within only a few picoseconds for $A$ \cite{Korn2011, Lagarde2014, Cadiz2016} and $B$ \cite{Palummo2015} complexes at low temperatures, it is likely that $A^-$ and $B^-$ trion formation is slow enough to compete with the population decay of the trion unbound states on comparable timescales in our CVD-grown sample. This is indeed in accord with measured $A^-$ trion formation times of $\sim 1$~ps in MoS$_2$ \cite{Lui2014} and $\sim 2$~ps in MoSe$_2$ \cite{Singh2016}. A quantitative model of the local doping concentration thus requires explicit knowledge of timescales for charge carrier capture and population decay not measured for our samples. At the qualitative level, the comparison of the maps in Figs.~\ref{fig4}(b) and \ref{fig4}(c) demonstrates that $A^-$ and $B^-$ trion photogeneration dynamics are not too distinct. In other words, $B$ exciton population relaxes on timescales comparable to their radiative decay, providing sufficient time for the formation of defect-localized $B^-$ trions. The reduced contrast and increased values of charged exciton densities in the doping profile of Fig.~\ref{fig4}(c) indicate for our sample that the ratio between the timescales of trion formation and neutral exciton decay ($\tau_\mathrm{f} / \tau$) is smaller for the $B$ manifold as compared to the $A$ manifold. This observation is most pronounced for $B^-$ trions in regular regions with low contamination.

In summary, spatially inhomogeneous distributions of charged excitons due to unintentional chemical doping were studied on CVD-grown MoS$_2$ MLs with PL imaging and spectroscopy. In addition to $A^-$ trions, our PL experiments identified defect-localized $B^-$ trions with binding energy of $34$~meV. Our analysis of charge doping profiles constructed from relative contributions of charged and neutral exciton PL indicates that the formation of localized $B^-$ trions competes with their relaxation and decay on comparable timescales.

We thank P.~Altpeter and R.~Rath for assistance in the clean room. This research was funded by the European Research Council under the ERC grant agreement no.\@ 336749, the Volkswagen Foundation, the Deutsche Forschungsgemeinschaft (DFG) Cluster of Excellence Nanosystems Initiative Munich (NIM) with financial support from the Center for NanoScience (CeNS) and LMUinnovativ. A.D.M. and H.Y. acknowledge support from the Laboratory Directed Research and Development (LDRD) program and the Center for Integrated Nanotechnologies at LANL.

%%%%%%%%%%%%%%%%%%%%%%%%%%%%%%%%%%%%%% BIBLIOGRAPHY %%%%%%%%%%%%%%%%%%%%%%%%%%%%%%%%%%%%%%%

%%%%%%%%%%%%%%%%%%%%%%%%%%%%%%%%%%%%%%%%%%%%%%%%%%%%%%%%%%%%%%%%%%%%%%%%%%%%%%%%%%%%%%%%%%%


\begin{thebibliography}{58}
\expandafter\ifx\csname natexlab\endcsname\relax\def\natexlab#1{#1}\fi
\expandafter\ifx\csname bibnamefont\endcsname\relax
  \def\bibnamefont#1{#1}\fi
\expandafter\ifx\csname bibfnamefont\endcsname\relax
  \def\bibfnamefont#1{#1}\fi
\expandafter\ifx\csname citenamefont\endcsname\relax
  \def\citenamefont#1{#1}\fi
\expandafter\ifx\csname url\endcsname\relax
  \def\url#1{\texttt{#1}}\fi
\expandafter\ifx\csname urlprefix\endcsname\relax\def\urlprefix{URL }\fi
\providecommand{\bibinfo}[2]{#2}
\providecommand{\eprint}[2][]{\url{#2}}

\bibitem[{\citenamefont{He et~al.}(2014)\citenamefont{He, Kumar, Zhao, Wang,
  Mak, Zhao, and Shan}}]{He2014}
\bibinfo{author}{\bibfnamefont{K.}~\bibnamefont{He}},
  \bibinfo{author}{\bibfnamefont{N.}~\bibnamefont{Kumar}},
  \bibinfo{author}{\bibfnamefont{L.}~\bibnamefont{Zhao}},
  \bibinfo{author}{\bibfnamefont{Z.}~\bibnamefont{Wang}},
  \bibinfo{author}{\bibfnamefont{K.~F.} \bibnamefont{Mak}},
  \bibinfo{author}{\bibfnamefont{H.}~\bibnamefont{Zhao}}, \bibnamefont{and}
  \bibinfo{author}{\bibfnamefont{J.}~\bibnamefont{Shan}},
  \bibinfo{journal}{Phys. Rev. Lett.} \textbf{\bibinfo{volume}{113}},
  \bibinfo{pages}{026803} (\bibinfo{year}{2014}).

\bibitem[{\citenamefont{Chernikov et~al.}(2014)\citenamefont{Chernikov,
  Berkelbach, Hill, Rigosi, Li, Aslan, Reichman, Hybertsen, and
  Heinz}}]{Chernikov2014}
\bibinfo{author}{\bibfnamefont{A.}~\bibnamefont{Chernikov}},
  \bibinfo{author}{\bibfnamefont{T.~C.} \bibnamefont{Berkelbach}},
  \bibinfo{author}{\bibfnamefont{H.~M.} \bibnamefont{Hill}},
  \bibinfo{author}{\bibfnamefont{A.}~\bibnamefont{Rigosi}},
  \bibinfo{author}{\bibfnamefont{Y.}~\bibnamefont{Li}},
  \bibinfo{author}{\bibfnamefont{O.~B.} \bibnamefont{Aslan}},
  \bibinfo{author}{\bibfnamefont{D.~R.} \bibnamefont{Reichman}},
  \bibinfo{author}{\bibfnamefont{M.~S.} \bibnamefont{Hybertsen}},
  \bibnamefont{and} \bibinfo{author}{\bibfnamefont{T.~F.} \bibnamefont{Heinz}},
  \bibinfo{journal}{Phys. Rev. Lett.} \textbf{\bibinfo{volume}{113}},
  \bibinfo{pages}{076802} (\bibinfo{year}{2014}).

\bibitem[{\citenamefont{Ye et~al.}(2014)\citenamefont{Ye, Cao, O'Brien, Zhu,
  Yin, Wang, Louie, and Zhang}}]{Ye2014}
\bibinfo{author}{\bibfnamefont{Z.}~\bibnamefont{Ye}},
  \bibinfo{author}{\bibfnamefont{T.}~\bibnamefont{Cao}},
  \bibinfo{author}{\bibfnamefont{K.}~\bibnamefont{O'Brien}},
  \bibinfo{author}{\bibfnamefont{H.}~\bibnamefont{Zhu}},
  \bibinfo{author}{\bibfnamefont{X.}~\bibnamefont{Yin}},
  \bibinfo{author}{\bibfnamefont{Y.}~\bibnamefont{Wang}},
  \bibinfo{author}{\bibfnamefont{S.~G.} \bibnamefont{Louie}}, \bibnamefont{and}
  \bibinfo{author}{\bibfnamefont{X.}~\bibnamefont{Zhang}},
  \bibinfo{journal}{Nature} \textbf{\bibinfo{volume}{513}},
  \bibinfo{pages}{214} (\bibinfo{year}{2014}).

\bibitem[{\citenamefont{Splendiani et~al.}(2010)\citenamefont{Splendiani, Sun,
  Zhang, Li, Kim, Chim, Galli, and Wang}}]{Splendiani2010}
\bibinfo{author}{\bibfnamefont{A.}~\bibnamefont{Splendiani}},
  \bibinfo{author}{\bibfnamefont{L.}~\bibnamefont{Sun}},
  \bibinfo{author}{\bibfnamefont{Y.}~\bibnamefont{Zhang}},
  \bibinfo{author}{\bibfnamefont{T.}~\bibnamefont{Li}},
  \bibinfo{author}{\bibfnamefont{J.}~\bibnamefont{Kim}},
  \bibinfo{author}{\bibfnamefont{C.-Y.} \bibnamefont{Chim}},
  \bibinfo{author}{\bibfnamefont{G.}~\bibnamefont{Galli}}, \bibnamefont{and}
  \bibinfo{author}{\bibfnamefont{F.}~\bibnamefont{Wang}},
  \bibinfo{journal}{Nano Lett.} \textbf{\bibinfo{volume}{10}},
  \bibinfo{pages}{1271} (\bibinfo{year}{2010}).

\bibitem[{\citenamefont{Mak et~al.}(2010)\citenamefont{Mak, Lee, Hone, Shan,
  and Heinz}}]{Mak2010}
\bibinfo{author}{\bibfnamefont{K.~F.} \bibnamefont{Mak}},
  \bibinfo{author}{\bibfnamefont{C.}~\bibnamefont{Lee}},
  \bibinfo{author}{\bibfnamefont{J.}~\bibnamefont{Hone}},
  \bibinfo{author}{\bibfnamefont{J.}~\bibnamefont{Shan}}, \bibnamefont{and}
  \bibinfo{author}{\bibfnamefont{T.~F.} \bibnamefont{Heinz}},
  \bibinfo{journal}{Phys. Rev. Lett.} \textbf{\bibinfo{volume}{105}},
  \bibinfo{pages}{136805} (\bibinfo{year}{2010}).

\bibitem[{\citenamefont{Mak et~al.}(2012)\citenamefont{Mak, He, Shan, and
  Heinz}}]{Mak2012}
\bibinfo{author}{\bibfnamefont{K.~F.} \bibnamefont{Mak}},
  \bibinfo{author}{\bibfnamefont{K.}~\bibnamefont{He}},
  \bibinfo{author}{\bibfnamefont{J.}~\bibnamefont{Shan}}, \bibnamefont{and}
  \bibinfo{author}{\bibfnamefont{T.~F.} \bibnamefont{Heinz}},
  \bibinfo{journal}{Nat. Nanotechnol.} \textbf{\bibinfo{volume}{7}},
  \bibinfo{pages}{494} (\bibinfo{year}{2012}).

\bibitem[{\citenamefont{Zeng et~al.}(2012)\citenamefont{Zeng, Dai, Yao, Xiao,
  and Cui}}]{Zeng2012}
\bibinfo{author}{\bibfnamefont{H.}~\bibnamefont{Zeng}},
  \bibinfo{author}{\bibfnamefont{J.}~\bibnamefont{Dai}},
  \bibinfo{author}{\bibfnamefont{W.}~\bibnamefont{Yao}},
  \bibinfo{author}{\bibfnamefont{D.}~\bibnamefont{Xiao}}, \bibnamefont{and}
  \bibinfo{author}{\bibfnamefont{X.}~\bibnamefont{Cui}}, \bibinfo{journal}{Nat.
  Nanotechnol.} \textbf{\bibinfo{volume}{7}}, \bibinfo{pages}{490}
  (\bibinfo{year}{2012}).

\bibitem[{\citenamefont{Jones et~al.}(2013)\citenamefont{Jones, Yu, Ghimire,
  Wu, Aivazian, Ross, Zhao, Yan, Mandrus, Xiao et~al.}}]{Jones2013}
\bibinfo{author}{\bibfnamefont{A.~M.} \bibnamefont{Jones}},
  \bibinfo{author}{\bibfnamefont{H.}~\bibnamefont{Yu}},
  \bibinfo{author}{\bibfnamefont{N.~J.} \bibnamefont{Ghimire}},
  \bibinfo{author}{\bibfnamefont{S.}~\bibnamefont{Wu}},
  \bibinfo{author}{\bibfnamefont{G.}~\bibnamefont{Aivazian}},
  \bibinfo{author}{\bibfnamefont{J.~S.} \bibnamefont{Ross}},
  \bibinfo{author}{\bibfnamefont{B.}~\bibnamefont{Zhao}},
  \bibinfo{author}{\bibfnamefont{J.}~\bibnamefont{Yan}},
  \bibinfo{author}{\bibfnamefont{D.~G.} \bibnamefont{Mandrus}},
  \bibinfo{author}{\bibfnamefont{D.}~\bibnamefont{Xiao}}, \bibnamefont{et~al.},
  \bibinfo{journal}{Nat. Nanotechnol.} \textbf{\bibinfo{volume}{8}},
  \bibinfo{pages}{634} (\bibinfo{year}{2013}).

\bibitem[{\citenamefont{Xu et~al.}(2014)\citenamefont{Xu, Yao, Xiao, and
  Heinz}}]{Xu2014}
\bibinfo{author}{\bibfnamefont{X.}~\bibnamefont{Xu}},
  \bibinfo{author}{\bibfnamefont{W.}~\bibnamefont{Yao}},
  \bibinfo{author}{\bibfnamefont{D.}~\bibnamefont{Xiao}}, \bibnamefont{and}
  \bibinfo{author}{\bibfnamefont{T.~F.} \bibnamefont{Heinz}},
  \bibinfo{journal}{Nat. Phys.} \textbf{\bibinfo{volume}{10}},
  \bibinfo{pages}{343} (\bibinfo{year}{2014}).

\bibitem[{\citenamefont{Schaibley et~al.}(2016)\citenamefont{Schaibley, Yu,
  Clark, Rivera, Ross, Seyler, Yao, and Xu}}]{Schaibley2016}
\bibinfo{author}{\bibfnamefont{J.~R.} \bibnamefont{Schaibley}},
  \bibinfo{author}{\bibfnamefont{H.}~\bibnamefont{Yu}},
  \bibinfo{author}{\bibfnamefont{G.}~\bibnamefont{Clark}},
  \bibinfo{author}{\bibfnamefont{P.}~\bibnamefont{Rivera}},
  \bibinfo{author}{\bibfnamefont{J.~S.} \bibnamefont{Ross}},
  \bibinfo{author}{\bibfnamefont{K.~L.} \bibnamefont{Seyler}},
  \bibinfo{author}{\bibfnamefont{W.}~\bibnamefont{Yao}}, \bibnamefont{and}
  \bibinfo{author}{\bibfnamefont{X.}~\bibnamefont{Xu}}, \bibinfo{journal}{Nat.
  Rev. Mater.} \textbf{\bibinfo{volume}{1}}, \bibinfo{pages}{16055}
  (\bibinfo{year}{2016}).

\bibitem[{\citenamefont{Mak et~al.}(2013)\citenamefont{Mak, He, Lee, Lee, Hone,
  Heinz, and Shan}}]{Mak2013}
\bibinfo{author}{\bibfnamefont{K.~F.} \bibnamefont{Mak}},
  \bibinfo{author}{\bibfnamefont{K.}~\bibnamefont{He}},
  \bibinfo{author}{\bibfnamefont{C.}~\bibnamefont{Lee}},
  \bibinfo{author}{\bibfnamefont{G.~H.} \bibnamefont{Lee}},
  \bibinfo{author}{\bibfnamefont{J.}~\bibnamefont{Hone}},
  \bibinfo{author}{\bibfnamefont{T.~F.} \bibnamefont{Heinz}}, \bibnamefont{and}
  \bibinfo{author}{\bibfnamefont{J.}~\bibnamefont{Shan}},
  \bibinfo{journal}{Nat. Mater.} \textbf{\bibinfo{volume}{12}},
  \bibinfo{pages}{207} (\bibinfo{year}{2013}).

\bibitem[{\citenamefont{Ross et~al.}(2013)\citenamefont{Ross, Wu, Yu, Ghimire,
  Jones, Aivazian, Yan, Mandrus, Xiao, Yao et~al.}}]{Ross2013}
\bibinfo{author}{\bibfnamefont{J.~S.} \bibnamefont{Ross}},
  \bibinfo{author}{\bibfnamefont{S.}~\bibnamefont{Wu}},
  \bibinfo{author}{\bibfnamefont{H.}~\bibnamefont{Yu}},
  \bibinfo{author}{\bibfnamefont{N.~J.} \bibnamefont{Ghimire}},
  \bibinfo{author}{\bibfnamefont{A.~M.} \bibnamefont{Jones}},
  \bibinfo{author}{\bibfnamefont{G.}~\bibnamefont{Aivazian}},
  \bibinfo{author}{\bibfnamefont{J.}~\bibnamefont{Yan}},
  \bibinfo{author}{\bibfnamefont{D.~G.} \bibnamefont{Mandrus}},
  \bibinfo{author}{\bibfnamefont{D.}~\bibnamefont{Xiao}},
  \bibinfo{author}{\bibfnamefont{W.}~\bibnamefont{Yao}}, \bibnamefont{et~al.},
  \bibinfo{journal}{Nat. Commun.} \textbf{\bibinfo{volume}{4}},
  \bibinfo{pages}{1474} (\bibinfo{year}{2013}).

\bibitem[{\citenamefont{You et~al.}(2015)\citenamefont{You, Zhang, Berkelbach,
  Hybertsen, Reichman, and Heinz}}]{You2015}
\bibinfo{author}{\bibfnamefont{Y.}~\bibnamefont{You}},
  \bibinfo{author}{\bibfnamefont{X.-X.} \bibnamefont{Zhang}},
  \bibinfo{author}{\bibfnamefont{T.~C.} \bibnamefont{Berkelbach}},
  \bibinfo{author}{\bibfnamefont{M.~S.} \bibnamefont{Hybertsen}},
  \bibinfo{author}{\bibfnamefont{D.~R.} \bibnamefont{Reichman}},
  \bibnamefont{and} \bibinfo{author}{\bibfnamefont{T.~F.} \bibnamefont{Heinz}},
  \bibinfo{journal}{Nat. Phys.} \textbf{\bibinfo{volume}{11}},
  \bibinfo{pages}{477} (\bibinfo{year}{2015}).

\bibitem[{\citenamefont{Sie et~al.}(2015)\citenamefont{Sie, Frenzel, Lee, Kong,
  and Gedik}}]{Sie2015}
\bibinfo{author}{\bibfnamefont{E.~J.} \bibnamefont{Sie}},
  \bibinfo{author}{\bibfnamefont{A.~J.} \bibnamefont{Frenzel}},
  \bibinfo{author}{\bibfnamefont{Y.-H.} \bibnamefont{Lee}},
  \bibinfo{author}{\bibfnamefont{J.}~\bibnamefont{Kong}}, \bibnamefont{and}
  \bibinfo{author}{\bibfnamefont{N.}~\bibnamefont{Gedik}},
  \bibinfo{journal}{Phys. Rev. B} \textbf{\bibinfo{volume}{92}},
  \bibinfo{pages}{125417} (\bibinfo{year}{2015}).

\bibitem[{\citenamefont{Hao et~al.}(2017)\citenamefont{Hao, Specht, Nagler, Xu,
  Tran, Singh, Dass, Sch{\"u}ller, Korn, Richter et~al.}}]{Hao2017}
\bibinfo{author}{\bibfnamefont{K.}~\bibnamefont{Hao}},
  \bibinfo{author}{\bibfnamefont{J.~F.} \bibnamefont{Specht}},
  \bibinfo{author}{\bibfnamefont{P.}~\bibnamefont{Nagler}},
  \bibinfo{author}{\bibfnamefont{L.}~\bibnamefont{Xu}},
  \bibinfo{author}{\bibfnamefont{K.}~\bibnamefont{Tran}},
  \bibinfo{author}{\bibfnamefont{A.}~\bibnamefont{Singh}},
  \bibinfo{author}{\bibfnamefont{C.~K.} \bibnamefont{Dass}},
  \bibinfo{author}{\bibfnamefont{C.}~\bibnamefont{Sch{\"u}ller}},
  \bibinfo{author}{\bibfnamefont{T.}~\bibnamefont{Korn}},
  \bibinfo{author}{\bibfnamefont{M.}~\bibnamefont{Richter}},
  \bibnamefont{et~al.}, \bibinfo{journal}{Nat. Commun.}
  \textbf{\bibinfo{volume}{8}}, \bibinfo{pages}{15552} (\bibinfo{year}{2017}).

\bibitem[{\citenamefont{Sidler et~al.}(2017)\citenamefont{Sidler, Back, Cotlet,
  Srivastava, Fink, Kroner, Demler, and Imamoglu}}]{Sidler2017}
\bibinfo{author}{\bibfnamefont{M.}~\bibnamefont{Sidler}},
  \bibinfo{author}{\bibfnamefont{P.}~\bibnamefont{Back}},
  \bibinfo{author}{\bibfnamefont{O.}~\bibnamefont{Cotlet}},
  \bibinfo{author}{\bibfnamefont{A.}~\bibnamefont{Srivastava}},
  \bibinfo{author}{\bibfnamefont{T.}~\bibnamefont{Fink}},
  \bibinfo{author}{\bibfnamefont{M.}~\bibnamefont{Kroner}},
  \bibinfo{author}{\bibfnamefont{E.}~\bibnamefont{Demler}}, \bibnamefont{and}
  \bibinfo{author}{\bibfnamefont{A.}~\bibnamefont{Imamoglu}},
  \bibinfo{journal}{Nat. Phys.} \textbf{\bibinfo{volume}{13}},
  \bibinfo{pages}{255} (\bibinfo{year}{2017}).

\bibitem[{\citenamefont{Almand-Hunter et~al.}(2014)\citenamefont{Almand-Hunter,
  Li, Cundiff, Mootz, Kira, and Koch}}]{Almand-Hunter2014}
\bibinfo{author}{\bibfnamefont{A.~E.} \bibnamefont{Almand-Hunter}},
  \bibinfo{author}{\bibfnamefont{H.}~\bibnamefont{Li}},
  \bibinfo{author}{\bibfnamefont{S.~T.} \bibnamefont{Cundiff}},
  \bibinfo{author}{\bibfnamefont{M.}~\bibnamefont{Mootz}},
  \bibinfo{author}{\bibfnamefont{M.}~\bibnamefont{Kira}}, \bibnamefont{and}
  \bibinfo{author}{\bibfnamefont{S.~W.} \bibnamefont{Koch}},
  \bibinfo{journal}{Nature} \textbf{\bibinfo{volume}{506}},
  \bibinfo{pages}{471} (\bibinfo{year}{2014}).

\bibitem[{\citenamefont{Portella-Oberli
  et~al.}(2009)\citenamefont{Portella-Oberli, Berney, Kappei, Morier-Genoud,
  Szczytko, and Deveaud-Pl\'edran}}]{Portella-Oberli2009}
\bibinfo{author}{\bibfnamefont{M.~T.} \bibnamefont{Portella-Oberli}},
  \bibinfo{author}{\bibfnamefont{J.}~\bibnamefont{Berney}},
  \bibinfo{author}{\bibfnamefont{L.}~\bibnamefont{Kappei}},
  \bibinfo{author}{\bibfnamefont{F.}~\bibnamefont{Morier-Genoud}},
  \bibinfo{author}{\bibfnamefont{J.}~\bibnamefont{Szczytko}}, \bibnamefont{and}
  \bibinfo{author}{\bibfnamefont{B.}~\bibnamefont{Deveaud-Pl\'edran}},
  \bibinfo{journal}{Phys. Rev. Lett.} \textbf{\bibinfo{volume}{102}},
  \bibinfo{pages}{096402} (\bibinfo{year}{2009}).

\bibitem[{\citenamefont{Najmaei et~al.}(2013)\citenamefont{Najmaei, Liu, Zhou,
  Zou, Shi, Lei, Yakobson, Idrobo, Ajayan, and Lou}}]{Najmaei2013}
\bibinfo{author}{\bibfnamefont{S.}~\bibnamefont{Najmaei}},
  \bibinfo{author}{\bibfnamefont{Z.}~\bibnamefont{Liu}},
  \bibinfo{author}{\bibfnamefont{W.}~\bibnamefont{Zhou}},
  \bibinfo{author}{\bibfnamefont{X.}~\bibnamefont{Zou}},
  \bibinfo{author}{\bibfnamefont{G.}~\bibnamefont{Shi}},
  \bibinfo{author}{\bibfnamefont{S.}~\bibnamefont{Lei}},
  \bibinfo{author}{\bibfnamefont{B.~I.} \bibnamefont{Yakobson}},
  \bibinfo{author}{\bibfnamefont{J.-C.} \bibnamefont{Idrobo}},
  \bibinfo{author}{\bibfnamefont{P.~M.} \bibnamefont{Ajayan}},
  \bibnamefont{and} \bibinfo{author}{\bibfnamefont{J.}~\bibnamefont{Lou}},
  \bibinfo{journal}{Nat. Mater.} \textbf{\bibinfo{volume}{12}},
  \bibinfo{pages}{754} (\bibinfo{year}{2013}).

\bibitem[{\citenamefont{Neumann et~al.}(2017)\citenamefont{Neumann, Lindlau,
  Colombier, Nutz, Najmaei, Lou, Mohite, Yamaguchi, and
  H{\"o}gele}}]{Neumann2017}
\bibinfo{author}{\bibfnamefont{A.}~\bibnamefont{Neumann}},
  \bibinfo{author}{\bibfnamefont{J.}~\bibnamefont{Lindlau}},
  \bibinfo{author}{\bibfnamefont{L.}~\bibnamefont{Colombier}},
  \bibinfo{author}{\bibfnamefont{M.}~\bibnamefont{Nutz}},
  \bibinfo{author}{\bibfnamefont{S.}~\bibnamefont{Najmaei}},
  \bibinfo{author}{\bibfnamefont{J.}~\bibnamefont{Lou}},
  \bibinfo{author}{\bibfnamefont{A.~D.} \bibnamefont{Mohite}},
  \bibinfo{author}{\bibfnamefont{H.}~\bibnamefont{Yamaguchi}},
  \bibnamefont{and}
  \bibinfo{author}{\bibfnamefont{A.}~\bibnamefont{H{\"o}gele}},
  \bibinfo{journal}{Nat. Nanotechnol.} \textbf{\bibinfo{volume}{12}},
  \bibinfo{pages}{329} (\bibinfo{year}{2017}).

\bibitem[{\citenamefont{Hong et~al.}(2015)\citenamefont{Hong, Hu, Probert, Li,
  Lv, Yang, Gu, Mao, Feng, Xie et~al.}}]{Hong2015}
\bibinfo{author}{\bibfnamefont{J.}~\bibnamefont{Hong}},
  \bibinfo{author}{\bibfnamefont{Z.}~\bibnamefont{Hu}},
  \bibinfo{author}{\bibfnamefont{M.}~\bibnamefont{Probert}},
  \bibinfo{author}{\bibfnamefont{K.}~\bibnamefont{Li}},
  \bibinfo{author}{\bibfnamefont{D.}~\bibnamefont{Lv}},
  \bibinfo{author}{\bibfnamefont{X.}~\bibnamefont{Yang}},
  \bibinfo{author}{\bibfnamefont{L.}~\bibnamefont{Gu}},
  \bibinfo{author}{\bibfnamefont{N.}~\bibnamefont{Mao}},
  \bibinfo{author}{\bibfnamefont{Q.}~\bibnamefont{Feng}},
  \bibinfo{author}{\bibfnamefont{L.}~\bibnamefont{Xie}}, \bibnamefont{et~al.},
  \bibinfo{journal}{Nat. Commun.} \textbf{\bibinfo{volume}{6}},
  \bibinfo{pages}{6293} (\bibinfo{year}{2015}).

\bibitem[{\citenamefont{Lin et~al.}(2016)\citenamefont{Lin, Carvalho, Kahn, Lv,
  Rao, Terrones, Pimenta, and Terrones}}]{Lin2016}
\bibinfo{author}{\bibfnamefont{Z.}~\bibnamefont{Lin}},
  \bibinfo{author}{\bibfnamefont{B.~R.} \bibnamefont{Carvalho}},
  \bibinfo{author}{\bibfnamefont{E.}~\bibnamefont{Kahn}},
  \bibinfo{author}{\bibfnamefont{R.}~\bibnamefont{Lv}},
  \bibinfo{author}{\bibfnamefont{R.}~\bibnamefont{Rao}},
  \bibinfo{author}{\bibfnamefont{H.}~\bibnamefont{Terrones}},
  \bibinfo{author}{\bibfnamefont{M.~A.} \bibnamefont{Pimenta}},
  \bibnamefont{and} \bibinfo{author}{\bibfnamefont{M.}~\bibnamefont{Terrones}},
  \bibinfo{journal}{2D Mater.} \textbf{\bibinfo{volume}{3}},
  \bibinfo{pages}{022002} (\bibinfo{year}{2016}).

\bibitem[{\citenamefont{Cai et~al.}(2016)\citenamefont{Cai, Zhou, Zhang, and
  Zhang}}]{Cai2016}
\bibinfo{author}{\bibfnamefont{Y.}~\bibnamefont{Cai}},
  \bibinfo{author}{\bibfnamefont{H.}~\bibnamefont{Zhou}},
  \bibinfo{author}{\bibfnamefont{G.}~\bibnamefont{Zhang}}, \bibnamefont{and}
  \bibinfo{author}{\bibfnamefont{Y.-W.} \bibnamefont{Zhang}},
  \bibinfo{journal}{Chem. Mater.} \textbf{\bibinfo{volume}{28}},
  \bibinfo{pages}{8611} (\bibinfo{year}{2016}).

\bibitem[{\citenamefont{Radisavljevic et~al.}(2011)\citenamefont{Radisavljevic,
  Radenovic, Brivio, Giacometti, and Kis}}]{Radisavljevic2011}
\bibinfo{author}{\bibfnamefont{B.}~\bibnamefont{Radisavljevic}},
  \bibinfo{author}{\bibfnamefont{A.}~\bibnamefont{Radenovic}},
  \bibinfo{author}{\bibfnamefont{J.}~\bibnamefont{Brivio}},
  \bibinfo{author}{\bibfnamefont{V.}~\bibnamefont{Giacometti}},
  \bibnamefont{and} \bibinfo{author}{\bibfnamefont{A.}~\bibnamefont{Kis}},
  \bibinfo{journal}{Nat. Nanotechnol.} \textbf{\bibinfo{volume}{6}},
  \bibinfo{pages}{147} (\bibinfo{year}{2011}).

\bibitem[{\citenamefont{Tongay et~al.}(2013{\natexlab{a}})\citenamefont{Tongay,
  Zhou, Ataca, Liu, Kang, Matthews, You, Li, Grossman, and Wu}}]{Tongay2013a}
\bibinfo{author}{\bibfnamefont{S.}~\bibnamefont{Tongay}},
  \bibinfo{author}{\bibfnamefont{J.}~\bibnamefont{Zhou}},
  \bibinfo{author}{\bibfnamefont{C.}~\bibnamefont{Ataca}},
  \bibinfo{author}{\bibfnamefont{J.}~\bibnamefont{Liu}},
  \bibinfo{author}{\bibfnamefont{J.~S.} \bibnamefont{Kang}},
  \bibinfo{author}{\bibfnamefont{T.~S.} \bibnamefont{Matthews}},
  \bibinfo{author}{\bibfnamefont{L.}~\bibnamefont{You}},
  \bibinfo{author}{\bibfnamefont{J.}~\bibnamefont{Li}},
  \bibinfo{author}{\bibfnamefont{J.~C.} \bibnamefont{Grossman}},
  \bibnamefont{and} \bibinfo{author}{\bibfnamefont{J.}~\bibnamefont{Wu}},
  \bibinfo{journal}{Nano Lett.} \textbf{\bibinfo{volume}{13}},
  \bibinfo{pages}{2831} (\bibinfo{year}{2013}{\natexlab{a}}).

\bibitem[{\citenamefont{Mouri et~al.}(2013)\citenamefont{Mouri, Miyauchi, and
  Matsuda}}]{Mouri2013}
\bibinfo{author}{\bibfnamefont{S.}~\bibnamefont{Mouri}},
  \bibinfo{author}{\bibfnamefont{Y.}~\bibnamefont{Miyauchi}}, \bibnamefont{and}
  \bibinfo{author}{\bibfnamefont{K.}~\bibnamefont{Matsuda}},
  \bibinfo{journal}{Nano Lett.} \textbf{\bibinfo{volume}{13}},
  \bibinfo{pages}{5944} (\bibinfo{year}{2013}).

\bibitem[{\citenamefont{Cadiz et~al.}(2016)\citenamefont{Cadiz, Tricard, Gay,
  Lagarde, Wang, Robert, Renucci, Urbaszek, and Marie}}]{Cadiz2016}
\bibinfo{author}{\bibfnamefont{F.}~\bibnamefont{Cadiz}},
  \bibinfo{author}{\bibfnamefont{S.}~\bibnamefont{Tricard}},
  \bibinfo{author}{\bibfnamefont{M.}~\bibnamefont{Gay}},
  \bibinfo{author}{\bibfnamefont{D.}~\bibnamefont{Lagarde}},
  \bibinfo{author}{\bibfnamefont{G.}~\bibnamefont{Wang}},
  \bibinfo{author}{\bibfnamefont{C.}~\bibnamefont{Robert}},
  \bibinfo{author}{\bibfnamefont{P.}~\bibnamefont{Renucci}},
  \bibinfo{author}{\bibfnamefont{B.}~\bibnamefont{Urbaszek}}, \bibnamefont{and}
  \bibinfo{author}{\bibfnamefont{X.}~\bibnamefont{Marie}},
  \bibinfo{journal}{Appl. Phys. Lett.} \textbf{\bibinfo{volume}{108}},
  \bibinfo{pages}{251106} (\bibinfo{year}{2016}).

\bibitem[{\citenamefont{Efimkin and MacDonald}(2017)}]{Efimkin2017}
\bibinfo{author}{\bibfnamefont{D.~K.} \bibnamefont{Efimkin}} \bibnamefont{and}
  \bibinfo{author}{\bibfnamefont{A.~H.} \bibnamefont{MacDonald}},
  \bibinfo{journal}{Phys. Rev. B} \textbf{\bibinfo{volume}{95}},
  \bibinfo{pages}{035417} (\bibinfo{year}{2017}).

\bibitem[{\citenamefont{Reina et~al.}(2008)\citenamefont{Reina, Son, Jiao, Fan,
  Dresselhaus, Liu, and Kong}}]{Reina2008}
\bibinfo{author}{\bibfnamefont{A.}~\bibnamefont{Reina}},
  \bibinfo{author}{\bibfnamefont{H.}~\bibnamefont{Son}},
  \bibinfo{author}{\bibfnamefont{L.}~\bibnamefont{Jiao}},
  \bibinfo{author}{\bibfnamefont{B.}~\bibnamefont{Fan}},
  \bibinfo{author}{\bibfnamefont{M.~S.} \bibnamefont{Dresselhaus}},
  \bibinfo{author}{\bibfnamefont{Z.}~\bibnamefont{Liu}}, \bibnamefont{and}
  \bibinfo{author}{\bibfnamefont{J.}~\bibnamefont{Kong}}, \bibinfo{journal}{J.
  Phys. Chem. C} \textbf{\bibinfo{volume}{112}}, \bibinfo{pages}{17741}
  (\bibinfo{year}{2008}).

\bibitem[{\citenamefont{Buscema et~al.}(2013)\citenamefont{Buscema, Barkelid,
  Zwiller, van~der Zant, Steele, and Castellanos-Gomez}}]{Buscema2013}
\bibinfo{author}{\bibfnamefont{M.}~\bibnamefont{Buscema}},
  \bibinfo{author}{\bibfnamefont{M.}~\bibnamefont{Barkelid}},
  \bibinfo{author}{\bibfnamefont{V.}~\bibnamefont{Zwiller}},
  \bibinfo{author}{\bibfnamefont{H.~S.~J.} \bibnamefont{van~der Zant}},
  \bibinfo{author}{\bibfnamefont{G.~A.} \bibnamefont{Steele}},
  \bibnamefont{and}
  \bibinfo{author}{\bibfnamefont{A.}~\bibnamefont{Castellanos-Gomez}},
  \bibinfo{journal}{Nano Lett.} \textbf{\bibinfo{volume}{13}},
  \bibinfo{pages}{358} (\bibinfo{year}{2013}).

\bibitem[{\citenamefont{Yamaguchi et~al.}(2015)\citenamefont{Yamaguchi,
  Blancon, Kappera, Lei, Najmaei, Mangum, Gupta, Ajayan, Lou, Chhowalla
  et~al.}}]{Yamaguchi2015}
\bibinfo{author}{\bibfnamefont{H.}~\bibnamefont{Yamaguchi}},
  \bibinfo{author}{\bibfnamefont{J.-C.} \bibnamefont{Blancon}},
  \bibinfo{author}{\bibfnamefont{R.}~\bibnamefont{Kappera}},
  \bibinfo{author}{\bibfnamefont{S.}~\bibnamefont{Lei}},
  \bibinfo{author}{\bibfnamefont{S.}~\bibnamefont{Najmaei}},
  \bibinfo{author}{\bibfnamefont{B.~D.} \bibnamefont{Mangum}},
  \bibinfo{author}{\bibfnamefont{G.}~\bibnamefont{Gupta}},
  \bibinfo{author}{\bibfnamefont{P.~M.} \bibnamefont{Ajayan}},
  \bibinfo{author}{\bibfnamefont{J.}~\bibnamefont{Lou}},
  \bibinfo{author}{\bibfnamefont{M.}~\bibnamefont{Chhowalla}},
  \bibnamefont{et~al.}, \bibinfo{journal}{ACS Nano}
  \textbf{\bibinfo{volume}{9}}, \bibinfo{pages}{840} (\bibinfo{year}{2015}).

\bibitem[{\citenamefont{Lee et~al.}(2014)\citenamefont{Lee, Lee, van~der Zande,
  Chen, Li, Han, Cui, Arefe, Nuckolls, Heinz et~al.}}]{Lee2014}
\bibinfo{author}{\bibfnamefont{C.-H.} \bibnamefont{Lee}},
  \bibinfo{author}{\bibfnamefont{G.-H.} \bibnamefont{Lee}},
  \bibinfo{author}{\bibfnamefont{A.~M.} \bibnamefont{van~der Zande}},
  \bibinfo{author}{\bibfnamefont{W.}~\bibnamefont{Chen}},
  \bibinfo{author}{\bibfnamefont{Y.}~\bibnamefont{Li}},
  \bibinfo{author}{\bibfnamefont{M.}~\bibnamefont{Han}},
  \bibinfo{author}{\bibfnamefont{X.}~\bibnamefont{Cui}},
  \bibinfo{author}{\bibfnamefont{G.}~\bibnamefont{Arefe}},
  \bibinfo{author}{\bibfnamefont{C.}~\bibnamefont{Nuckolls}},
  \bibinfo{author}{\bibfnamefont{T.~F.} \bibnamefont{Heinz}},
  \bibnamefont{et~al.}, \bibinfo{journal}{Nat. Nanotechnol.}
  \textbf{\bibinfo{volume}{9}}, \bibinfo{pages}{676} (\bibinfo{year}{2014}).

\bibitem[{\citenamefont{Dolui et~al.}(2013)\citenamefont{Dolui, Rungger, and
  Sanvito}}]{Dolui2013}
\bibinfo{author}{\bibfnamefont{K.}~\bibnamefont{Dolui}},
  \bibinfo{author}{\bibfnamefont{I.}~\bibnamefont{Rungger}}, \bibnamefont{and}
  \bibinfo{author}{\bibfnamefont{S.}~\bibnamefont{Sanvito}},
  \bibinfo{journal}{Phys. Rev. B} \textbf{\bibinfo{volume}{87}},
  \bibinfo{pages}{165402} (\bibinfo{year}{2013}).

\bibitem[{\citenamefont{Scheuschner et~al.}(2014)\citenamefont{Scheuschner,
  Ochedowski, Kaulitz, Gillen, Schleberger, and Maultzsch}}]{Scheuschner2014}
\bibinfo{author}{\bibfnamefont{N.}~\bibnamefont{Scheuschner}},
  \bibinfo{author}{\bibfnamefont{O.}~\bibnamefont{Ochedowski}},
  \bibinfo{author}{\bibfnamefont{A.-M.} \bibnamefont{Kaulitz}},
  \bibinfo{author}{\bibfnamefont{R.}~\bibnamefont{Gillen}},
  \bibinfo{author}{\bibfnamefont{M.}~\bibnamefont{Schleberger}},
  \bibnamefont{and}
  \bibinfo{author}{\bibfnamefont{J.}~\bibnamefont{Maultzsch}},
  \bibinfo{journal}{Phys. Rev. B} \textbf{\bibinfo{volume}{89}},
  \bibinfo{pages}{125406} (\bibinfo{year}{2014}).

\bibitem[{\citenamefont{Kang and Han}(2017)}]{Kang2017}
\bibinfo{author}{\bibfnamefont{Y.}~\bibnamefont{Kang}} \bibnamefont{and}
  \bibinfo{author}{\bibfnamefont{S.}~\bibnamefont{Han}},
  \bibinfo{journal}{Nanoscale} \textbf{\bibinfo{volume}{9}},
  \bibinfo{pages}{4265} (\bibinfo{year}{2017}).

\bibitem[{\citenamefont{Peimyoo et~al.}(2014)\citenamefont{Peimyoo, Yang,
  Shang, Shen, Wang, and Yu}}]{Peimyoo2014}
\bibinfo{author}{\bibfnamefont{N.}~\bibnamefont{Peimyoo}},
  \bibinfo{author}{\bibfnamefont{W.}~\bibnamefont{Yang}},
  \bibinfo{author}{\bibfnamefont{J.}~\bibnamefont{Shang}},
  \bibinfo{author}{\bibfnamefont{X.}~\bibnamefont{Shen}},
  \bibinfo{author}{\bibfnamefont{Y.}~\bibnamefont{Wang}}, \bibnamefont{and}
  \bibinfo{author}{\bibfnamefont{T.}~\bibnamefont{Yu}}, \bibinfo{journal}{ACS
  Nano} \textbf{\bibinfo{volume}{8}}, \bibinfo{pages}{11320}
  (\bibinfo{year}{2014}).

\bibitem[{\citenamefont{Lui et~al.}(2014)\citenamefont{Lui, Frenzel, Pilon,
  Lee, Ling, Akselrod, Kong, and Gedik}}]{Lui2014}
\bibinfo{author}{\bibfnamefont{C.~H.} \bibnamefont{Lui}},
  \bibinfo{author}{\bibfnamefont{A.~J.} \bibnamefont{Frenzel}},
  \bibinfo{author}{\bibfnamefont{D.~V.} \bibnamefont{Pilon}},
  \bibinfo{author}{\bibfnamefont{Y.-H.} \bibnamefont{Lee}},
  \bibinfo{author}{\bibfnamefont{X.}~\bibnamefont{Ling}},
  \bibinfo{author}{\bibfnamefont{G.~M.} \bibnamefont{Akselrod}},
  \bibinfo{author}{\bibfnamefont{J.}~\bibnamefont{Kong}}, \bibnamefont{and}
  \bibinfo{author}{\bibfnamefont{N.}~\bibnamefont{Gedik}},
  \bibinfo{journal}{Phys. Rev. Lett.} \textbf{\bibinfo{volume}{113}},
  \bibinfo{pages}{166801} (\bibinfo{year}{2014}).

\bibitem[{\citenamefont{Tongay et~al.}(2013{\natexlab{b}})\citenamefont{Tongay,
  Suh, Ataca, Fan, Luce, Kang, Liu, Ko, Raghunathanan, Zhou
  et~al.}}]{Tongay2013}
\bibinfo{author}{\bibfnamefont{S.}~\bibnamefont{Tongay}},
  \bibinfo{author}{\bibfnamefont{J.}~\bibnamefont{Suh}},
  \bibinfo{author}{\bibfnamefont{C.}~\bibnamefont{Ataca}},
  \bibinfo{author}{\bibfnamefont{W.}~\bibnamefont{Fan}},
  \bibinfo{author}{\bibfnamefont{A.}~\bibnamefont{Luce}},
  \bibinfo{author}{\bibfnamefont{J.~S.} \bibnamefont{Kang}},
  \bibinfo{author}{\bibfnamefont{J.}~\bibnamefont{Liu}},
  \bibinfo{author}{\bibfnamefont{C.}~\bibnamefont{Ko}},
  \bibinfo{author}{\bibfnamefont{R.}~\bibnamefont{Raghunathanan}},
  \bibinfo{author}{\bibfnamefont{J.}~\bibnamefont{Zhou}}, \bibnamefont{et~al.},
  \bibinfo{journal}{Sci. Rep.} \textbf{\bibinfo{volume}{3}},
  \bibinfo{pages}{2657} (\bibinfo{year}{2013}{\natexlab{b}}).

\bibitem[{\citenamefont{Nan et~al.}(2014)\citenamefont{Nan, Wang, Wang, Liang,
  Lu, Chen, He, Tan, Miao, Wang et~al.}}]{Nan2014}
\bibinfo{author}{\bibfnamefont{H.}~\bibnamefont{Nan}},
  \bibinfo{author}{\bibfnamefont{Z.}~\bibnamefont{Wang}},
  \bibinfo{author}{\bibfnamefont{W.}~\bibnamefont{Wang}},
  \bibinfo{author}{\bibfnamefont{Z.}~\bibnamefont{Liang}},
  \bibinfo{author}{\bibfnamefont{Y.}~\bibnamefont{Lu}},
  \bibinfo{author}{\bibfnamefont{Q.}~\bibnamefont{Chen}},
  \bibinfo{author}{\bibfnamefont{D.}~\bibnamefont{He}},
  \bibinfo{author}{\bibfnamefont{P.}~\bibnamefont{Tan}},
  \bibinfo{author}{\bibfnamefont{F.}~\bibnamefont{Miao}},
  \bibinfo{author}{\bibfnamefont{X.}~\bibnamefont{Wang}}, \bibnamefont{et~al.},
  \bibinfo{journal}{ACS Nano} \textbf{\bibinfo{volume}{8}},
  \bibinfo{pages}{5738} (\bibinfo{year}{2014}).

\bibitem[{\citenamefont{Godde et~al.}(2016)\citenamefont{Godde, Schmidt,
  Schmutzler, A\ss{}mann, Debus, Withers, Alexeev, Del Pozo-Zamudio, Skrypka,
  Novoselov et~al.}}]{Godde2016}
\bibinfo{author}{\bibfnamefont{T.}~\bibnamefont{Godde}},
  \bibinfo{author}{\bibfnamefont{D.}~\bibnamefont{Schmidt}},
  \bibinfo{author}{\bibfnamefont{J.}~\bibnamefont{Schmutzler}},
  \bibinfo{author}{\bibfnamefont{M.}~\bibnamefont{A\ss{}mann}},
  \bibinfo{author}{\bibfnamefont{J.}~\bibnamefont{Debus}},
  \bibinfo{author}{\bibfnamefont{F.}~\bibnamefont{Withers}},
  \bibinfo{author}{\bibfnamefont{E.~M.} \bibnamefont{Alexeev}},
  \bibinfo{author}{\bibfnamefont{O.}~\bibnamefont{Del Pozo-Zamudio}},
  \bibinfo{author}{\bibfnamefont{O.~V.} \bibnamefont{Skrypka}},
  \bibinfo{author}{\bibfnamefont{K.~S.} \bibnamefont{Novoselov}},
  \bibnamefont{et~al.}, \bibinfo{journal}{Phys. Rev. B}
  \textbf{\bibinfo{volume}{94}}, \bibinfo{pages}{165301}
  (\bibinfo{year}{2016}).

\bibitem[{\citenamefont{Berkelbach et~al.}(2013)\citenamefont{Berkelbach,
  Hybertsen, and Reichman}}]{Berkelbach2013}
\bibinfo{author}{\bibfnamefont{T.~C.} \bibnamefont{Berkelbach}},
  \bibinfo{author}{\bibfnamefont{M.~S.} \bibnamefont{Hybertsen}},
  \bibnamefont{and} \bibinfo{author}{\bibfnamefont{D.~R.}
  \bibnamefont{Reichman}}, \bibinfo{journal}{Phys. Rev. B}
  \textbf{\bibinfo{volume}{88}}, \bibinfo{pages}{045318}
  (\bibinfo{year}{2013}).

\bibitem[{\citenamefont{Chernikov et~al.}(2015)\citenamefont{Chernikov, van~der
  Zande, Hill, Rigosi, Velauthapillai, Hone, and Heinz}}]{Chernikov2015a}
\bibinfo{author}{\bibfnamefont{A.}~\bibnamefont{Chernikov}},
  \bibinfo{author}{\bibfnamefont{A.~M.} \bibnamefont{van~der Zande}},
  \bibinfo{author}{\bibfnamefont{H.~M.} \bibnamefont{Hill}},
  \bibinfo{author}{\bibfnamefont{A.~F.} \bibnamefont{Rigosi}},
  \bibinfo{author}{\bibfnamefont{A.}~\bibnamefont{Velauthapillai}},
  \bibinfo{author}{\bibfnamefont{J.}~\bibnamefont{Hone}}, \bibnamefont{and}
  \bibinfo{author}{\bibfnamefont{T.~F.} \bibnamefont{Heinz}},
  \bibinfo{journal}{Phys. Rev. Lett.} \textbf{\bibinfo{volume}{115}},
  \bibinfo{pages}{126802} (\bibinfo{year}{2015}).

\bibitem[{\citenamefont{Courtade et~al.}(2017)\citenamefont{Courtade, Semina,
  Manca, Glazov, Robert, Cadiz, Wang, Taniguchi, Watanabe, Pierre
  et~al.}}]{Courtade2017}
\bibinfo{author}{\bibfnamefont{E.}~\bibnamefont{Courtade}},
  \bibinfo{author}{\bibfnamefont{M.}~\bibnamefont{Semina}},
  \bibinfo{author}{\bibfnamefont{M.}~\bibnamefont{Manca}},
  \bibinfo{author}{\bibfnamefont{M.~M.} \bibnamefont{Glazov}},
  \bibinfo{author}{\bibfnamefont{C.}~\bibnamefont{Robert}},
  \bibinfo{author}{\bibfnamefont{F.}~\bibnamefont{Cadiz}},
  \bibinfo{author}{\bibfnamefont{G.}~\bibnamefont{Wang}},
  \bibinfo{author}{\bibfnamefont{T.}~\bibnamefont{Taniguchi}},
  \bibinfo{author}{\bibfnamefont{K.}~\bibnamefont{Watanabe}},
  \bibinfo{author}{\bibfnamefont{M.}~\bibnamefont{Pierre}},
  \bibnamefont{et~al.}, \bibinfo{journal}{Phys. Rev. B}
  \textbf{\bibinfo{volume}{96}}, \bibinfo{pages}{085302}
  (\bibinfo{year}{2017}).

\bibitem[{\citenamefont{Xiao et~al.}(2012)\citenamefont{Xiao, Liu, Feng, Xu,
  and Yao}}]{Xiao2012}
\bibinfo{author}{\bibfnamefont{D.}~\bibnamefont{Xiao}},
  \bibinfo{author}{\bibfnamefont{G.-B.} \bibnamefont{Liu}},
  \bibinfo{author}{\bibfnamefont{W.}~\bibnamefont{Feng}},
  \bibinfo{author}{\bibfnamefont{X.}~\bibnamefont{Xu}}, \bibnamefont{and}
  \bibinfo{author}{\bibfnamefont{W.}~\bibnamefont{Yao}},
  \bibinfo{journal}{Phys. Rev. Lett.} \textbf{\bibinfo{volume}{108}},
  \bibinfo{pages}{196802} (\bibinfo{year}{2012}).

\bibitem[{\citenamefont{Stier et~al.}(2016)\citenamefont{Stier, McCreary,
  Jonker, Kono, and Crooker}}]{Stier2016}
\bibinfo{author}{\bibfnamefont{A.~V.} \bibnamefont{Stier}},
  \bibinfo{author}{\bibfnamefont{K.~M.} \bibnamefont{McCreary}},
  \bibinfo{author}{\bibfnamefont{B.~T.} \bibnamefont{Jonker}},
  \bibinfo{author}{\bibfnamefont{J.}~\bibnamefont{Kono}}, \bibnamefont{and}
  \bibinfo{author}{\bibfnamefont{S.~A.} \bibnamefont{Crooker}},
  \bibinfo{journal}{Nat. Commun.} \textbf{\bibinfo{volume}{7}},
  \bibinfo{pages}{10643} (\bibinfo{year}{2016}).

\bibitem[{\citenamefont{Borzda et~al.}(2015)\citenamefont{Borzda, Gadermaier,
  Vujicic, Topolovsek, Borovsak, Mertelj, Viola, Manzoni, Pogna, Brida
  et~al.}}]{Borzda2014}
\bibinfo{author}{\bibfnamefont{T.}~\bibnamefont{Borzda}},
  \bibinfo{author}{\bibfnamefont{C.}~\bibnamefont{Gadermaier}},
  \bibinfo{author}{\bibfnamefont{N.}~\bibnamefont{Vujicic}},
  \bibinfo{author}{\bibfnamefont{P.}~\bibnamefont{Topolovsek}},
  \bibinfo{author}{\bibfnamefont{M.}~\bibnamefont{Borovsak}},
  \bibinfo{author}{\bibfnamefont{T.}~\bibnamefont{Mertelj}},
  \bibinfo{author}{\bibfnamefont{D.}~\bibnamefont{Viola}},
  \bibinfo{author}{\bibfnamefont{C.}~\bibnamefont{Manzoni}},
  \bibinfo{author}{\bibfnamefont{E.~A.~A.} \bibnamefont{Pogna}},
  \bibinfo{author}{\bibfnamefont{D.}~\bibnamefont{Brida}},
  \bibnamefont{et~al.}, \bibinfo{journal}{Adv. Funct. Mater.}
  \textbf{\bibinfo{volume}{25}}, \bibinfo{pages}{3351} (\bibinfo{year}{2015}).

\bibitem[{\citenamefont{He et~al.}(2013)\citenamefont{He, Poole, Mak, and
  Shan}}]{He2013}
\bibinfo{author}{\bibfnamefont{K.}~\bibnamefont{He}},
  \bibinfo{author}{\bibfnamefont{C.}~\bibnamefont{Poole}},
  \bibinfo{author}{\bibfnamefont{K.~F.} \bibnamefont{Mak}}, \bibnamefont{and}
  \bibinfo{author}{\bibfnamefont{J.}~\bibnamefont{Shan}},
  \bibinfo{journal}{Nano Lett.} \textbf{\bibinfo{volume}{13}},
  \bibinfo{pages}{2931} (\bibinfo{year}{2013}).

\bibitem[{\citenamefont{Conley et~al.}(2013)\citenamefont{Conley, Wang,
  Ziegler, Haglund, Pantelides, and Bolotin}}]{Conley2013}
\bibinfo{author}{\bibfnamefont{H.~J.} \bibnamefont{Conley}},
  \bibinfo{author}{\bibfnamefont{B.}~\bibnamefont{Wang}},
  \bibinfo{author}{\bibfnamefont{J.~I.} \bibnamefont{Ziegler}},
  \bibinfo{author}{\bibfnamefont{R.~F.} \bibnamefont{Haglund}},
  \bibinfo{author}{\bibfnamefont{S.~T.} \bibnamefont{Pantelides}},
  \bibnamefont{and} \bibinfo{author}{\bibfnamefont{K.~I.}
  \bibnamefont{Bolotin}}, \bibinfo{journal}{Nano Lett.}
  \textbf{\bibinfo{volume}{13}}, \bibinfo{pages}{3626} (\bibinfo{year}{2013}).

\bibitem[{\citenamefont{Zhu et~al.}(2013)\citenamefont{Zhu, Wang, Liu, Marie,
  Qiao, Zhang, Wu, Fan, Tan, Amand et~al.}}]{Zhu2013}
\bibinfo{author}{\bibfnamefont{C.~R.} \bibnamefont{Zhu}},
  \bibinfo{author}{\bibfnamefont{G.}~\bibnamefont{Wang}},
  \bibinfo{author}{\bibfnamefont{B.~L.} \bibnamefont{Liu}},
  \bibinfo{author}{\bibfnamefont{X.}~\bibnamefont{Marie}},
  \bibinfo{author}{\bibfnamefont{X.~F.} \bibnamefont{Qiao}},
  \bibinfo{author}{\bibfnamefont{X.}~\bibnamefont{Zhang}},
  \bibinfo{author}{\bibfnamefont{X.~X.} \bibnamefont{Wu}},
  \bibinfo{author}{\bibfnamefont{H.}~\bibnamefont{Fan}},
  \bibinfo{author}{\bibfnamefont{P.~H.} \bibnamefont{Tan}},
  \bibinfo{author}{\bibfnamefont{T.}~\bibnamefont{Amand}},
  \bibnamefont{et~al.}, \bibinfo{journal}{Phys. Rev. B}
  \textbf{\bibinfo{volume}{88}}, \bibinfo{pages}{121301}
  (\bibinfo{year}{2013}).

\bibitem[{\citenamefont{Rice et~al.}(2013)\citenamefont{Rice, Young, Zan,
  Bangert, Wolverson, Georgiou, Jalil, and Novoselov}}]{Rice2013}
\bibinfo{author}{\bibfnamefont{C.}~\bibnamefont{Rice}},
  \bibinfo{author}{\bibfnamefont{R.~J.} \bibnamefont{Young}},
  \bibinfo{author}{\bibfnamefont{R.}~\bibnamefont{Zan}},
  \bibinfo{author}{\bibfnamefont{U.}~\bibnamefont{Bangert}},
  \bibinfo{author}{\bibfnamefont{D.}~\bibnamefont{Wolverson}},
  \bibinfo{author}{\bibfnamefont{T.}~\bibnamefont{Georgiou}},
  \bibinfo{author}{\bibfnamefont{R.}~\bibnamefont{Jalil}}, \bibnamefont{and}
  \bibinfo{author}{\bibfnamefont{K.~S.} \bibnamefont{Novoselov}},
  \bibinfo{journal}{Phys. Rev. B} \textbf{\bibinfo{volume}{87}},
  \bibinfo{pages}{081307} (\bibinfo{year}{2013}).

\bibitem[{\citenamefont{Chakraborty et~al.}(2012)\citenamefont{Chakraborty,
  Bera, Muthu, Bhowmick, Waghmare, and Sood}}]{Chakraborty2012}
\bibinfo{author}{\bibfnamefont{B.}~\bibnamefont{Chakraborty}},
  \bibinfo{author}{\bibfnamefont{A.}~\bibnamefont{Bera}},
  \bibinfo{author}{\bibfnamefont{D.~V.~S.} \bibnamefont{Muthu}},
  \bibinfo{author}{\bibfnamefont{S.}~\bibnamefont{Bhowmick}},
  \bibinfo{author}{\bibfnamefont{U.~V.} \bibnamefont{Waghmare}},
  \bibnamefont{and} \bibinfo{author}{\bibfnamefont{A.~K.} \bibnamefont{Sood}},
  \bibinfo{journal}{Phys. Rev. B} \textbf{\bibinfo{volume}{85}},
  \bibinfo{pages}{161403} (\bibinfo{year}{2012}).

\bibitem[{\citenamefont{Lanzillo et~al.}(2013)\citenamefont{Lanzillo,
  Glen~Birdwell, Amani, Crowne, Shah, Najmaei, Liu, Ajayan, Lou, Dubey
  et~al.}}]{Lanzillo2013}
\bibinfo{author}{\bibfnamefont{N.~A.} \bibnamefont{Lanzillo}},
  \bibinfo{author}{\bibfnamefont{A.}~\bibnamefont{Glen~Birdwell}},
  \bibinfo{author}{\bibfnamefont{M.}~\bibnamefont{Amani}},
  \bibinfo{author}{\bibfnamefont{F.~J.} \bibnamefont{Crowne}},
  \bibinfo{author}{\bibfnamefont{P.~B.} \bibnamefont{Shah}},
  \bibinfo{author}{\bibfnamefont{S.}~\bibnamefont{Najmaei}},
  \bibinfo{author}{\bibfnamefont{Z.}~\bibnamefont{Liu}},
  \bibinfo{author}{\bibfnamefont{P.~M.} \bibnamefont{Ajayan}},
  \bibinfo{author}{\bibfnamefont{J.}~\bibnamefont{Lou}},
  \bibinfo{author}{\bibfnamefont{M.}~\bibnamefont{Dubey}},
  \bibnamefont{et~al.}, \bibinfo{journal}{Appl. Phys. Lett.}
  \textbf{\bibinfo{volume}{103}}, \bibinfo{pages}{093102}
  (\bibinfo{year}{2013}).

\bibitem[{\citenamefont{Qiu et~al.}(2013)\citenamefont{Qiu, Xu, Wang, Ren, Nan,
  Ni, Chen, Yuan, Miao, Song et~al.}}]{Qiu2013}
\bibinfo{author}{\bibfnamefont{H.}~\bibnamefont{Qiu}},
  \bibinfo{author}{\bibfnamefont{T.}~\bibnamefont{Xu}},
  \bibinfo{author}{\bibfnamefont{Z.}~\bibnamefont{Wang}},
  \bibinfo{author}{\bibfnamefont{W.}~\bibnamefont{Ren}},
  \bibinfo{author}{\bibfnamefont{H.}~\bibnamefont{Nan}},
  \bibinfo{author}{\bibfnamefont{Z.}~\bibnamefont{Ni}},
  \bibinfo{author}{\bibfnamefont{Q.}~\bibnamefont{Chen}},
  \bibinfo{author}{\bibfnamefont{S.}~\bibnamefont{Yuan}},
  \bibinfo{author}{\bibfnamefont{F.}~\bibnamefont{Miao}},
  \bibinfo{author}{\bibfnamefont{F.}~\bibnamefont{Song}}, \bibnamefont{et~al.},
  \bibinfo{journal}{Nat. Commun.} \textbf{\bibinfo{volume}{4}},
  \bibinfo{pages}{2642} (\bibinfo{year}{2013}).

\bibitem[{\citenamefont{Zhou et~al.}(2013)\citenamefont{Zhou, Zou, Najmaei,
  Liu, Shi, Kong, Lou, Ajayan, Yakobson, and Idrobo}}]{Zhou2013}
\bibinfo{author}{\bibfnamefont{W.}~\bibnamefont{Zhou}},
  \bibinfo{author}{\bibfnamefont{X.}~\bibnamefont{Zou}},
  \bibinfo{author}{\bibfnamefont{S.}~\bibnamefont{Najmaei}},
  \bibinfo{author}{\bibfnamefont{Z.}~\bibnamefont{Liu}},
  \bibinfo{author}{\bibfnamefont{Y.}~\bibnamefont{Shi}},
  \bibinfo{author}{\bibfnamefont{J.}~\bibnamefont{Kong}},
  \bibinfo{author}{\bibfnamefont{J.}~\bibnamefont{Lou}},
  \bibinfo{author}{\bibfnamefont{P.~M.} \bibnamefont{Ajayan}},
  \bibinfo{author}{\bibfnamefont{B.~I.} \bibnamefont{Yakobson}},
  \bibnamefont{and} \bibinfo{author}{\bibfnamefont{J.-C.}
  \bibnamefont{Idrobo}}, \bibinfo{journal}{Nano Lett.}
  \textbf{\bibinfo{volume}{13}}, \bibinfo{pages}{2615} (\bibinfo{year}{2013}).

\bibitem[{\citenamefont{Korn et~al.}(2011)\citenamefont{Korn, Heydrich, Hirmer,
  Schmutzler, and Sch{\"u}l\-ler}}]{Korn2011}
\bibinfo{author}{\bibfnamefont{T.}~\bibnamefont{Korn}},
  \bibinfo{author}{\bibfnamefont{S.}~\bibnamefont{Heydrich}},
  \bibinfo{author}{\bibfnamefont{M.}~\bibnamefont{Hirmer}},
  \bibinfo{author}{\bibfnamefont{J.}~\bibnamefont{Schmutzler}},
  \bibnamefont{and}
  \bibinfo{author}{\bibfnamefont{C.}~\bibnamefont{Sch{\"u}l\-ler}},
  \bibinfo{journal}{Appl. Phys. Lett.} \textbf{\bibinfo{volume}{99}},
  \bibinfo{pages}{102109} (\bibinfo{year}{2011}).

\bibitem[{\citenamefont{Lagarde et~al.}(2014)\citenamefont{Lagarde, Bouet,
  Marie, Zhu, Liu, Amand, Tan, and Urbaszek}}]{Lagarde2014}
\bibinfo{author}{\bibfnamefont{D.}~\bibnamefont{Lagarde}},
  \bibinfo{author}{\bibfnamefont{L.}~\bibnamefont{Bouet}},
  \bibinfo{author}{\bibfnamefont{X.}~\bibnamefont{Marie}},
  \bibinfo{author}{\bibfnamefont{C.~R.} \bibnamefont{Zhu}},
  \bibinfo{author}{\bibfnamefont{B.~L.} \bibnamefont{Liu}},
  \bibinfo{author}{\bibfnamefont{T.}~\bibnamefont{Amand}},
  \bibinfo{author}{\bibfnamefont{P.~H.} \bibnamefont{Tan}}, \bibnamefont{and}
  \bibinfo{author}{\bibfnamefont{B.}~\bibnamefont{Urbaszek}},
  \bibinfo{journal}{Phys. Rev. Lett.} \textbf{\bibinfo{volume}{112}},
  \bibinfo{pages}{047401} (\bibinfo{year}{2014}).

\bibitem[{\citenamefont{Palummo et~al.}(2015)\citenamefont{Palummo, Bernardi,
  and Grossman}}]{Palummo2015}
\bibinfo{author}{\bibfnamefont{M.}~\bibnamefont{Palummo}},
  \bibinfo{author}{\bibfnamefont{M.}~\bibnamefont{Bernardi}}, \bibnamefont{and}
  \bibinfo{author}{\bibfnamefont{J.~C.} \bibnamefont{Grossman}},
  \bibinfo{journal}{Nano Lett.} \textbf{\bibinfo{volume}{15}},
  \bibinfo{pages}{2794} (\bibinfo{year}{2015}).

\bibitem[{\citenamefont{Singh et~al.}(2016)\citenamefont{Singh, Moody, Tran,
  Scott, Overbeck, Bergh\"auser, Schaibley, Seifert, Pleskot, Gabor
  et~al.}}]{Singh2016}
\bibinfo{author}{\bibfnamefont{A.}~\bibnamefont{Singh}},
  \bibinfo{author}{\bibfnamefont{G.}~\bibnamefont{Moody}},
  \bibinfo{author}{\bibfnamefont{K.}~\bibnamefont{Tran}},
  \bibinfo{author}{\bibfnamefont{M.~E.} \bibnamefont{Scott}},
  \bibinfo{author}{\bibfnamefont{V.}~\bibnamefont{Overbeck}},
  \bibinfo{author}{\bibfnamefont{G.}~\bibnamefont{Bergh\"auser}},
  \bibinfo{author}{\bibfnamefont{J.}~\bibnamefont{Schaibley}},
  \bibinfo{author}{\bibfnamefont{E.~J.} \bibnamefont{Seifert}},
  \bibinfo{author}{\bibfnamefont{D.}~\bibnamefont{Pleskot}},
  \bibinfo{author}{\bibfnamefont{N.~M.} \bibnamefont{Gabor}},
  \bibnamefont{et~al.}, \bibinfo{journal}{Phys. Rev. B}
  \textbf{\bibinfo{volume}{93}}, \bibinfo{pages}{041401}
  (\bibinfo{year}{2016}).

\end{thebibliography}
\end{document}